\documentclass[dvips]{arxstspdf}
\usepackage{graphicx}
\usepackage{flushend,stfloats}

\volume{25}
\issue{3}
\pubyear{2010}
\firstpage{408}
\lastpage{428}
\doi{10.1214/09-STS292}

\begin{document}
\begin{frontmatter}

\title{A Conversation with George C. Tiao}
\runtitle{A Conversation with George C. Tiao}

\begin{aug}
\author[a]{\fnms{Daniel} \snm{Pe\~{n}a}\ead[label=e1]{daniel.pena@uc3m.es}}
\and
\author[b]{\fnms{Ruey S.} \snm{Tsay}\corref{}\ead[label=e2]{ruey.tsay@chicagobooth.edu}}
\runauthor{D. Pe\~{n}a and R. S. Tsay}

\affiliation{Universidad Carlos III de Madrid and University of
Chicago}

\address[a]{Daniel Pe\~{n}a is Professor and Rector, Departamento de Estadistica,
Universidad Carlos III de Madrid, Madrid, Spain
(\printead{e1}).}
\address[b]{Ruey S. Tsay is H.G.B. Alexander Professor of Econometrics and Statistics, Booth School of
Business, University of Chicago, 5807 S. Noodlawn Avenue, Chicago,
Illinois, USA (\printead{e2}).}

\end{aug}

\begin{abstract}
George C. Tiao was born in London in 1933. After graduating with a B.A.
in Economics from National Taiwan University in 1955 he went to the US
to obtain an M.B.A from New York University in 1958 and a Ph.D. in
Economics from the University of Wisconsin, Madison in 1962. From 1962
to 1982 he was Assistant, Associate, Professor and Bascom Professor of
Statistics and Business at the University of Wisconsin, Madison, and in
the period 1973--1975 was Chairman of the Department of Statistics. He
moved to the Graduate School of Business at the University of Chicago in
1982 and is the W. Allen Wallis Professor of Econometrics and Statistics
(\textit{emeritus}).

George Tiao has played a leading role in the development of Bayesian
Statistics, Time Series Analysis and Environmental Statistics. He is
co-author, with G.E.P. Box, of \textit{Bayesian Inference in Statistical
Analysis} and is the developer of a model-based approach to seasonal
adjustment (with S. C. Hillmer), of outlier analysis in time series
(with I. Chang), and of new ways of vector ARMA model building (with
R. S. Tsay). He is the author/co-author/co-editor of 7 books and over 120
articles in refereed econometric, environmental and statistical journals
and has been thesis advisor of over 25 students. He is a leading figure
in the development of Statistics in Taiwan and China and is the Founding
President of the International Chinese Statistical Association 1987--1988
and the Founding Chair Editor of the journal \textit{Statistica Sinica}
1988--1993. He played a leading role (over the 20 year period 1979--1999)
in the organization of the annual NBER/NSF Time Series Workshop and he
was a founding member of the annual conference ``Making Statistics More
Effective in Schools of Business'' 1986--2006. Among other honors he was
elected ASA Fellow (1973), IMS Fellow (1974), member of Academia Sinica,
Taiwan (1976) and ISI (1980), and was recipient of the Distinguished
Service Medal, DGBAS, Taiwan 1993, the Julius Shiskin Award, 2001, the
Wilks Memorial Medal Award, 2001, and the Statistician of the Year Award
in 2005 (ASA Chicago Chapter). He received honorary doctorates in 2003
from the Universidad Carlos III de Madrid and National Tsinghua
University, Hsinchu, Taiwan.

\end{abstract}


\end{frontmatter}

The original conversation took place in May 2003 in Chicago. Due to
George's health problems right after he retired in 2003, this final
version was finished in~2009.

Q: George, tell us something about your parents.

GCT: My mother was from a somewhat well-to-do family in Kunming, Yunnan,
but her father died early and she was brought up by her mother. In the
mid-1920s she went to Beijing Female Normal University, which was a famous
university at that time, and she caught the revolutionary fervor and
joined the Nationalists Army doing political work. My father was from a
very poor family in Szechuan and had a very hard time until he finally
graduated from college. When the Nationalists established the government
in China, they had exams every year to send students abroad and my
parents passed those exams and in 1930 my father went to Harvard and my
mother went to Illinois at Champaign. Some of my parents' friends
introduced them when they were in the US and my father transferred
from Harvard to Champaign--Urbana and they got married there.

I think they were in this country for about three years before they went
to England to study at the LSE (London School of Economics). I was born
there in 1933. But shortly after I was born, they ran out of money and
went back to China, so I was in England for just about four months. They
never had in mind anything about birth certificates and the only
concrete evidence I had about my place of birth was a slip, a little
sort of invoice-receipt for five pounds, that they obtained when they
checked out of a hospital. My mother always kept that, and when I came
to this country she gave it to me. I kept that in my wallet and carried
it with me all the time. Unfortunately that wallet was stolen in Paris
and I lost it.

Q: What did your parents do when they returned to China?

GCT: They went back to Shanghai. My father joined the Central Bank of
China and my mother was teaching accounting in Shanghai and maybe was
doing some job for the Nationalist party organization. When the
Sino--Japanese war started in 1937 my father was sent to establish a
branch of the Central Bank in Chungking. The year the war ended (1945)
was the year I graduated from the grade school.

Q: When did your family move to Taiwan?

GCT: There were four or five years of complete chaos in China. During
the war my father became the treasury head of the city of Chungking and
my mother was the principal of a high school, the first female high
school principal in Chungking. The World War ended in 1945, but in China
the war continued between the Nationalists and the Communists. In 1949
the situation was very hard for the Nationalists and my father got the
job to move all the gold and foreign currency reserves from Chungking to
Taiwan. My father was very loyal and a clean civil servant and I have
been very proud of him all my life. I remember when we went to the bank
and they took all the money (silver and gold) that was left in the vault
in four or five trucks. Each truck had two army guards, and at four or
five o'clock in the morning, we went from the bank to the airport. The
normal travel time to the airport was about an hour, but it took us
about six or seven hours or maybe eight hours, because the traffic was a
complete jam. That was the first time I witnessed a city and an army
about to collapse.

We went to the airport and they loaded the money on a plane. We got on
the airplane and for a while we just stopped and waited for the plane to
take off. I peaked out from the window and saw a machine gun pointed at
the airplane and my father was arguing with the soldiers. He was jumping
up and down saying this is the government's money, you cannot take any
of this out. My father was an extremely loyal civil servant to the
government and was completely oblivious of the risk. Finally the local
garrison commander talked to us, and my father let him, I think, take
about two or three boxes of the silver dollars out. When you think about
it, most people would just do whatever they can to save their lives, but
my father was arguing with them, defending the government property, and
this incident has left a~very, very deep impression on me about public
service. From there we went to Hong Kong.

Q: At any time did you feel that your life was really at risk?

GCT: No, we didn't know that because I was, in a~way, stunned by the
whole event and just followed my parents. I didn't know about the
dangers. Only later on did we realize that, and there were rumors that
we got robbed and that the whole family had perished.

We went to Hong Kong and in January 1950 we flew from Hong Kong to
Taiwan. I remember all my parents' friends in Hong Kong said there's no
point for us to go there because the whole thing will be over in three
months. All the people with some means stayed and we were really lucky
because my parents had no property, so we had nothing to lose. Well,
nothing to live either, except with the government. That was the
situation I was brought up in. So we went to Taiwan and, of course,
stayed much longer than three months. In June~1950, the Korean War
started. I always remember when my father came home one day at noon. At
the time, he always came home for lunch, had some rest, and then went
back to work again. He said to my mother, ``Well, now we don't have to
jump into the ocean.'' The way he said it, it's the Chinese way of
saying it, is that, ``We will not be eaten by fish.''

\section{Arriving in the United States}

Q: When and why did you come to the United States?

GCT: I came here in 1956. I finished high school in 1951 and got in to
National Taiwan University in the fall. I graduated in 1955, spent a
year in the army and then I came to the States in October 1956. See, at
that time, most of the college students came to study here and I got
admitted to NYU's economics department. My father arranged something
very helpful to me and I became a trainee for two years at a bank in New
York. A trainee there simply meant that I had to go through the training
in the morning. Well, first of all, the bank gave me \$50 a week, and
this was quite a bit of money at the time. In the morning I spent four
hours in their different departments and in the evening I went to
school. I was originally admitted to the econ department, but once I was
in New York City I met some other friends in the Bank of China. Some
people said, ``Why do you go to NYU in Washington Square? It's far away
from here. Why don't you go to NYU's business school, which is across
the street from here? Also, if you get a business degree job
opportunities are probably better.'' That's how I got into the business
school, because of its location. I spent four hours in the morning going
through training in the bank, in the afternoon I stayed in the bank
library, and in the evening I went to school. So that was sort of what
my life was like in New York.

Q: How did you decide to study statistics?

GCT: Well, it was sort of by accident. I met Barbara in high school in
Taiwan. We were engaged there in 1955. I came out first, then she came
out and we got married in 1958, the same year I got the MBA. And then
our first daughter was born in 1959. It was very tough; very uncertain
and very tough. We knew that we had to get out of New York because at
that time the bank job was just \$75
a week. There was a senior guy, a Chinese fellow, in the bank, who was
in his forties. His name was Chiu and one day we were having lunch
together and he said, ``George, you better leave. There's no chance. You
make seventy-five, right? I make a hundred and twenty-five.''

Q: That's your future.

GCT: After fifteen years. Then he gave me another example. He says,
``I'll tell you about my\break brother-in law who happens to be Gregory Chow.
He just got a Ph.D. and is a famous sort of economist, a famous MIT
professor.'' He told me all about Gregory. He says, ``You definitely
should get a Ph.D. and leave the bank.'' So I took his advice. You know
what happened? After I left New York the banks started opening up. He
jumped to American Express. In 1970 when I went back to New York with my
father to visit the banks, the guy was a senior VP at American Express.
I told him, ``Look, bad advice. Otherwise I'd be a heck of a lot
richer.'' And we laughed.

So that's how I went to Wisconsin. I got a scholarship, \$1500 a year,
great. My idea was to get a Ph.D. in international finance and I needed
a secondary field. The first obvious choice was accounting, because I
had all these cost accounting, advanced accounting and so forth in NYU
and back in National Taiwan University. I~went to the accounting
department and said well maybe I could take one advanced course and then
take a seminar. That should satisfy the secondary field. They said, no,
you have to take all these basic courses. Because I~didn't want to take
them the third time, one guy suggested, how about statistics? I had a
course in statistics in my sophomore year and it's like Greek, you know.
I went to talk to the guy (who) taught statistics in the business school
and he was so surprised that a student wanted a secondary field in
statistics. This never happened before. He said, ``Well, we really don't
have any advanced courses for you for the secondary field. To use that
as a secondary field, you have to show some advance courses.'' Then he
said that he heard that the math department was starting out a theory
course and I went to the math department. In the first year I took from
the math department a math stat course, and the teacher was a complete
disaster. Sam Wu, who became a famous engineering statistician later on,
was also in the class. I ended up working with somebody in class and we
studied the Mood book together and we did every problem in the book. So
at the end of the course I became one of the most advanced students in
statistics on campus. At the beginning of the second year, I was
persuaded to go back to economics. I thought I should transfer to some
other place, like Michigan or Stanford, but then I was told that two
good guys were coming this year. One is Goldberger and the other is Box.
It was lucky that I stayed. And in the third year I shifted from
economics to statistics with the permission from economics to write a
thesis with Box on Bayesian robustness. To qualify it as an econometrics
thesis, Goldberger finally made me write a piece about estimating common
parameters from two regressions with different variances. I did it in
the Bayesian framework and he was happy with it. Later on I worked with
Arnold Zellner in this area. Arnold joined Wisconsin just right after I
graduated, he read my thesis and we started working together. A person
who helped me a lot was Marvin Zelen. At that time he was at the
Mathematics Research Center, MRC, as a visitor. He was always very
encouraging and helpful to me in learning statistics. When the Raiffa
and Schlaiffer book on Applied Statistical Decision Theory came out,
Marvin organized a group seminar on the book at MRC. I learned so much
with that group, talking about decision trees, posterior analysis and
all that stuff.

\section{Bayesian Statistics}

Q: How was the subject of your thesis chosen?

GCT: The title of my thesis was Bayesian Analysis of Statistical
Assumptions. Box was interested in two things when he came to Wisconsin
in 1959. One was time series and the other was Bayes. Box was 42 when he
went to Wisconsin and he was at the peak of his ideas at the time. I
think he was frustrated by the frequentist approach because you have to
have sufficiency otherwise it becomes very difficult. He started looking
at Bayes. At the time Savage had a little book on stable estimation and
Box was studying that and got very interested. I remember his first
lecture the first year that he was teaching. There were six students in
his class, among them Bill Hunter, Sam Wu and I. His first couple
lectures were on the likelihood function of nonlinear models and how to
combine that with locally uniform priors. Nobody knew what he was
talking about. In a matter of two weeks it was clear that we were going
nowhere with this. He probably did look at our faces and saw them. Then
he completely changed it. He said, ``OK, let us start from scratch.'' He
got C. R. Rao's first book, not the Linear Statistical Models. He
started to get material from that book and began with expected value and
then only toward the very end of the second semester he came back to
Bayes and stuff like that. But in the meantime he was very interested in
trying the robustness in the Bayesian way and this was my dissertation
topic.

Q: Is there anything that you remember that impressed you very much at
the time of the Bayesian approach?

GCT: By the time I finished the thesis in 1962, you could just see the
easy way that this approach can study robustness from a different point
of view. If you draw a conclusion from the data, then the likelihood is
the most natural thing. You don't have to compare that with the things
that could have happened. So, it's a new approach. You learn, for
example, in the variance component, that if the variance component has a
negative estimate, it becomes a very bothersome thing. But, if you're
doing the Bayesian approach, you don't have that problem. After I had
done the random effect model I became more and more convinced about the
Bayesian approach.

Q: When you got your Ph.D. degree did you consider going elsewhere
instead of staying in Wisconsin?

GCT: When I was about finished, I went to George and said that, well, I
probably should start looking for a job. And his response was ``why
don't you stay here?'' He wanted to have joint appointments with other
schools. He himself and Bill Hunter were with engineering, and then at
that time John Gurland started with the medical school. In his mind
definitely something had to be done with economics and so I was a
natural person to have a joint appointment. George probably didn't mind
much about using his own students, but I guess in the Economics
Department they were having some reluctance because not hiring your own
students was a very good American tradition. However, the Business
School agreed to pay forty percent of my salary and I didn't have to do
anything for them for the first three years. So the first three years at
Wisconsin I taught one course a semester. And George Box said, ``Well
George, why don't you teach my course?'' I was shaking. I didn't know
what to do, you know, because I~had so little training and knew so
little about anything. In addition to Rao, I remember the three books
that I~used, more like a self-study really. The first is Kendall and
Stuart, it has a vast coverage but the only trouble is that there were a
lot of mistakes in there. The second is Wilks' book. I think I used
Wilks' book quite a bit the second year in fixing up my notes. Then the
other one is this book by Fisk.

Q: What kind of a schedule did you have at that time? Did you keep
working with George Box?

GCT: There are several major co-workers. One is Irwin Guttman, we shared
an office for four and one-half years. Then there's Arnold Zellner.
Arnold actually spent about half of the time inside the Statistics
Department. He and I ended up writing three papers together. Irwin and I
probably had four or something like that. And then George Box, I
continued to work with him. In the 60s we worked mostly on Bayesian
stuff. The only time series we did together was the one which is the
precursor of the Intervention Analysis paper: a change in level in a
nonstationary time series. I~think I started teaching time series in the
late 60s.

Q: What course did you teach?

GCT: There were two time series courses. George was teaching one (701)
and I was teaching a baby Box and Jenkins course for Business and
economic students. When I developed the course, I was also working with
Howard Thompson on the telephone paper---that is how I really learned
the seasonal models and so forth by working on the telephone data. Also,
one famous guy who attended every lecture was Sam Wu and that's how he
later on got into time series. So by the time I~got to England I think I
got more and more interested in time series.

\begin{figure}

\includegraphics{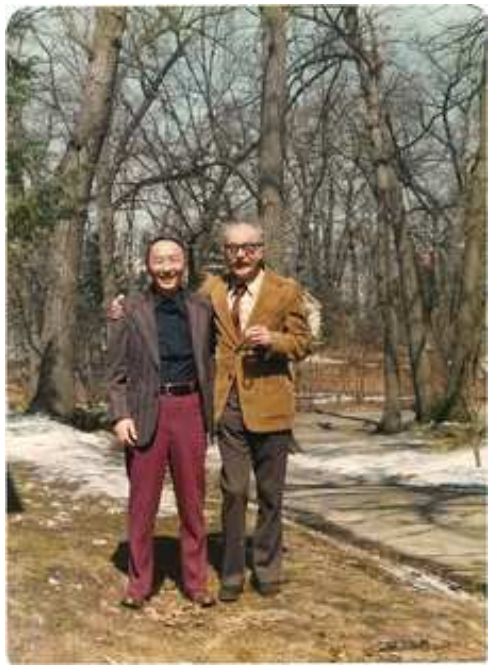}

\caption{With George Box in the 1960s.}
\end{figure}

Q: Did you have any trouble publishing the paper from your thesis?

GCT: No. I was not sure about the paper, but Box was quite sure and he
was right. I told him, ``Well there isn't much math in there,'' but the
ideas were very interesting where it worked out all this posterior
distribution and the different kind of robustness and so forth. George
was thinking this would get a good reception in \textit{Biometrika}, and sure we
did. We sent the paper in and Egon Pearson, who was the Editor, wrote a
long letter to Box saying that he thought the idea we had in the paper
was quite interesting. And he said that he and Neyman struggled a lot
trying to build statistics on Bayes, but in the end couldn't accept this
subjective idea and so forth. So they couldn't justify the prior
distribution because degree of belief is not frequency, and therefore he
gave up (laughing). Then he developed the well-known Neyman--Pearson
theory. It was quite sympathetic and very encouraging to us.

From then I was lucky, because I never had any trouble publishing
anything. The first time I had a little trouble publishing was in 1970,
so I was really very lucky the first eight years. Later I thought that
it was probably because the basic idea was quite new and to look at
problems in a Bayesian way was a very popular thing.

At that time one of the major things we (Box and I) were doing was
writing the Bayes book. We went to Harvard in 1965 to finish two books.
The first is the Bayes book and we also wanted to write a book on
Mathematical Statistics. But, of course, we ended up with half a book on
Bayes. The most interesting things I remember when working with George
in 1965 and 1966 in writing the book are data translated likelihood with
the noninformative priors, and the random effect model.

Now let me say something about both. We had this chapter, Chapter 2,
about Standard Normal Theory Inference Problems, and two people have
helped us a lot on this. One is Fred Mosteller and another is Jim
Dickey. They read many of the topics and said, well, we justify
everything on Jeffreys prior but nobody understands it. Why should you
hide behind Jeffreys? You should come out with arguments. It is by his
kind criticism that we went back and worked at it and then finally got a
likelihood function that when we take the transformation it becomes
normal. I think the data translated idea is really from Mosteller's
criticism and it makes the argument a lot more intuitive. So we spent
quite a lot of time on that subject.

The other thing we spent a lot of time on is the random effect model. A
bit of that was done before by W. Y. Tan and I, but we really worked
very hard on this problem in Boston. There are two aspects in the random
effect model. The first is looking at the estimation of variance
components. You have negative variance components estimates from random
effects models which are bothersome, and you also have them from mixed
and hierarchical models. But that is all on estimating variance
components.

The second and more interesting thing is the estimation of the random
effects because that's really kind of Stein's shrinkage and stuff like
that, and the two of us worked for a long, long time on this. If you
have many groups, you may assume that all the groups have the same mean,
just one mean. The second is the fixed-effect model that means that all
the groups means have locally uniform prior distributions. If you assume
they're totally spread out, then you don't give them a chance to shrink.
So the random effect model is very nice because it allows the group
means to come from some common population.

In this framework you invite the data to comment on the possibility of
clustering. So we thought that's really an interesting way, as we said
in Chapter 7 of the book. In fact, later on, all these things about
random effect hierarchical models in marketing and all that, this idea
of inviting the data to tell you whether you have clusters or not,
persisted. If you use fixed-effect models, you don't give them a chance.

The other thing we did at that time is this 1968 outlier paper which has
generated quite a bit of follow-ups. The only thing that was regretted
is that we didn't put the outlier paper in the book. But by then the
book was in the printing process in England so that we couldn't change
it any more.

Q: How was the situation of Bayes theory at that time?

GCT: The Neyman--Pearson approach is really formalized in their 1928
paper. So the basic foundations were only effective in the 30s. And the
next 10--20 years really developed that. All along the way there's
always this Bayesian stuff lurking in the background. The sequential
analysis was developed in the 40s. These are probably two of the main
things. There are always some doubts about this Neyman--Pearson framework
and in the 50s people started to think about alternative ways of drawing
inference. And there is also this likelihood function because Fisher
introduced the idea of likelihood function to obtain the maximum
likelihood estimates and so forth.

The likelihood function is a summary of information, and it has it's own
natural appeal. It's not just to get the maximum and then suddenly turn
around to find the sampling (frequentist) property of the estimates. So
there is talking about all this likelihood and thinking about
Bayes and, I think, Savages' idea about stable estimation and locally
uniform prior which produce an answer that looks the same as the
classical answer ``t'' and so forth, that has a tremendous impact on a lot
of people who are looking for alternatives. Take George Box as an
example. He was totally trained in sampling and all his early work on
robustness was of this frequentist type of thing. He was very frustrated
that in some cases, such as when you have sufficient statistics, the
answer to an inferential problem is very obvious. As soon as you go away
from sufficient statistics, how do you find the critical region and all
the similar regions, which is the basis of the hypothesis testing? So
that when you don't have sufficient statistics all this becomes
difficult. And when they become difficult people will naturally think
about alternatives and the likelihood thing becomes very hot in the 50s.
A lot of very good people like Barnard appeared and there is a famous
paper by Birnbaum. This all happened in the 50s, the late 50s and early
60s.

So there's a revival of the Bayesian framework or another critical look
at the frequentist approach. And, of course, the major problem at the
time, from our contention, is the Behrens--Fisher problem at the time.
Very often a locally uniform prior or something like that will produce a
Bayesian answer which is very similar, or which is the same thing, at
least in terms of practical use, to a sampling procedure. But
Behrens--Fisher, what is it? Lots of people tried very hard to understand
Fiducial and Bayesians got it, but it is very different from the
sampling approach. So that is the first clear-cut example of the
distinction between Bayesian and sampling approach.

So that was the time I became a student in statistics and read all this
stuff. And at the time I was kind of very depressed because I got out of
economics because there were too many theories on the one hand. On the
other hand, now you got into statistics and the first thing you jumped
into is the inference business. And I happened to have a major professor
interested in this and got me this problem. To work on this problem, I
started to read this literature and it was quite a turmoil at the time.

By the 60s people said, well, let's look at this. All these problems
that create difficulties in the frequentist approach and how do they
look if we adopt a Bayesian approach. At the beginning I think that's
reasonable because when you start a new theory, or revive a theory, the
first thing you always try to compare with is what the current
dominating theory has to say about this. And compare that answer with
the Bayesian answer. That, in a way, is the reason why we do all this. A
lot of the problems we are interested in is in that vain. And at that
time many papers are like that if you look at the papers by Geisser and
so forth; I just want to basically say a bit about the time. And that
was the reason why I~was interested in the variance component
estimation.

The variance component topic is very confusing and not well taught at
all. If every group has an equal number of observations, at least you
get sufficient statistics and inference is easy. As soon as the numbers
of observations of each group are different, things become very messy.
Apart from this, there is the possibility of negative variance
estimates, which is very counter-intuitive and people regard it as a
thorny problem. How can you have an approach, which works very nicely
only under certain sorts of nice balanced designs and immediately
becomes very difficult and you can't explain? As another example, take
the problem of comparing two means. If you teach a student how to tell
if the variances are equal, we can make inference. If the variances are
not equal, suddenly we cannot make any inference. That is the
Behrens--Fisher problem. In the analysis of variance, if the numbers of
observations in each group are equal, things are very simple. But as
soon as a number of groups are unbalanced, immediately they become very
difficult. For practitioners, if they really want to understand that,
they would ask why do you have a theory like that? That is all the
reasons why we want to look at all these things from a new point of
view, the Bayesian point of view. In the Bayesian approach, if you look
at all these thorny problems, you will produce good answers. And again,
at the same time, you always want to see in the simple cases you will
give similar answers. I don't know whether I explained the sentiment or
not, but I think it's a new insight comparison. You have to compare with
the dominating approach. Any time you want to make any changes you
cannot just throw them away. They are there. Most of the people believe
in them and so you have to defend [them] and say, Hey! We produce
answers as good as you are in the case where you can solve the problem,
but in the case where you cannot solve the problem we get answers, which
seem to be intuitively reasonable and asymptotically the same. Also, we
can get finite sample solutions. So you can remember Arnold Zellner,
even down to this day, always talks about finite sample solutions and
that's exactly the kind of thing that people like us talked about in the
60s. Well, he is the person from that generation anyway and is always
very proud that Bayesian can produce finite-sample solutions to all
these problems for any sample size. I can produce a posterior
distribution. The only difficulty is the numerical integration. I don't
have to find sufficient statistics. I don't have to find similar
regions.

Q: Why do you think that these ideas were not more widely accepted in
the statistical profession? Is it because of the computational
difficulty?

GCT: Well, there are many answers to that. One is, of course, there is
still a lot of frequentist people who think that the whole prior idea is
wrong. Everything has to be based on frequency and so forth. That's one
thing.

And the other thing is this. To some Bayesians, at least I believe that
in the 60s, all the problems are solved; the only thing that's left is
computation, and the computation is difficult at that time. You can do
numerical integration in one- or two-dimensional cases. So you talk to
Bayesians, all the problems are solved. You have got to assess the
prior. Once you assess the prior, you get the likelihood and bang!

In the 60s many frequentists thought that the\break Bayesians were basically
just hanging around. But somehow the attitude gradually shifted in the
70s because they found that the Bayes is a way to get good estimates.
So then they began to think that Bayes is probably not that bad at all.

\section{Time Series}

Q: Was it at the end of the 60s that you started to get involved in time
series analysis?

GCT: Yeah, more and more. Two things stood out for me in the beginning
of the 70s. One is this work with Bill Cleveland about the X-11. And that
took a very long time. The other is the level change paper for
nonstationary time series in 1996.

Q: But how you do become interested in seasonal adjustment?

GCT: Well, I was lecturing the Box and Jenkins method at Dupont doing
all this seasonal stuff. It was natural for me because I did the
telephone paper with Thompson. I had the feeling that I really had some
feeling for that kind of model. So I was particularly enthusiastic about
that. There was a guy, I forget his name now, who was in the audience
and he was an important statistician working there in the head office
and doing a lot of economic analysis. He took me to his office and said,
``Well, all of what you said sounds very interesting, but we've often
used the X-11 program from the Census Bureau.'' (We used various filters
and so forth. At that time I didn't even know what X-11 is.) And then he
said, ``Can you tell me about how these two things are related?'' So I
said, ``Well, I really know nothing about it, but I'll look.''

I came back and started to look into this and found filtering very
fascinating, so I started to play with it myself. Then came Bill
Cleveland. He was looking for a thesis topic and had an Electrical
Engineering background, so he was very familiar with spectral analysis
and the filters. We started to work together on that. A~rough logic goes
like this. Given a filter people use, we tried to figure out if the
filter has any kind of model background. In other words, is there any
possibility of modeling those filters? This is just like the way you can
think about ``exponential smoothing.'' People used it for many years,
but it's only later on that Muth came out with an explanation that the
exponential smoothing filter really produces the optimal forecast with
respect to the first-order integrated moving average model. By this
logic, we can bring together statistical modeling and choice of filters.

I kept thinking about a model-based explanation of the filters because I
always had the exponential smoothing case appearing in my head. And so
that's how Bill and I got started on that and we finally found some
approximate model for the Census filters. Harry Roberts and Arnold
Zellner became very high on the subject. It was the first time that
somebody found an empirically developed filter that became the widely
used official method, with a model-based explanation. So we published
the paper. We started working on that in 1969. But the final paper was
published in 1976.

Q: Yes, it took a long time.

GCT: Yeah, from 1969 to 1976. It took a long time to understand
everything. Going through it, I have a much better feeling about filters
because you have to deal with stationary and nonstationary stuff. For
all this, the traditional filtering doesn't work. I mean, you had to
have a theory behind it, so we worked out a theory and so forth. Later
on somebody says others have similar ideas, but when we were writing our
paper we didn't even know.

This was a very good learning process with Bill to get that out. Also,
about the time that paper was published, I was on the Advisory Committee
of Bureau of the Census and the Bureau had two conferences about
seasonal adjustment. Arnold Zellner was the leader. And we all got
involved. All those things together got me really interested in the
seasonal adjustment methods. Steve Hillmer, following Bill, was working
with me and so we continue to work on the problem. This is the part now
where I got into seasonal adjustment.

Q: You also wrote the intervention paper, and it has become a classic
article.

GCT: I don't know whether I deserve anything like that. Very often the
stuff that you got a lot of publicity over is not the thing that you
spent much time with. What happened was that after the 1966 paper,
people in education and psychology picked up the work. I got a lot of
phone calls from the education/psychology colleagues at Wisconsin. And
then there's somebody outside of Wisconsin who came and talked to me.
They applied the method because I guess it is much easier to understand
a change in the level in practice because there is no mean in a
nonstationary series. A change in mean is, of course, a very common
method. But here there is no mean and you weight the observations with
more weights for those observations close to the point of change and
then less and less. It's very intuitive, and education/psychology people
picked it up. I think the first paper they had was an application using
the Connecticut turnpike. There was a speed limit, but a lot of people
were still speeding, so police had a crackdown. They wanted to know what
is the effect of this crackdown in changing the level of the speed
violation, so it is a natural application. And they found that the
solutions that George and I worked out are much more intuitive than the
one they used before. So they applied our method to that kind of a
problem. In fact, this guy McGuire coined the word ``intervention
analysis.'' It's not George's or my invention. And he came and talked to
me.

There was another interesting application. At the turn of the century
Germany passed a new divorce law and they wanted to know how that
divorce law affects the divorce rate and the reconciliation rate. These
people and I actually published a paper on this. I even got some money
out of it (laughing). The paper was published in the \textit{Northwestern Law
Review}, which is not a bad law journal. Then people came and asked me
(and George) about application in advertisements. In advertisement
you'll see that the effect of advertising wears out in a cycle. A lot of
applications are of this type.

In 1974 or 1975 \textit{JASA} invited us to write a paper. We felt we should cover
all these consulting questions and put them all together. So we wrote
this intervention analysis paper for \textit{JASA}. I think putting the whole
thing together enabled us to come out with all these dynamic models.
This work only took about a week or two because we already knew all the
materials.

When we were about to submit, I told George that this is a bit thin and
we perhaps should do a bit more, like working out the algebra and
implications of all these different filters and different interventions
for the patterns. In the process of working it out, I say, ``Hey, this
can also be used for outlier testing or for missing observations.'' So
for the second part of the intervention paper I remember I spent a
couple of weeks doing algebra like crazy. Now the paper has more
substance (laughing). I don't know how many people paid attention to the
latter part, but actually the latter part later on has a lot of impact
on all the things I deal with, such as outliers, level shifts and stuff
like that. If you look at the second half of the paper, it's all there
somehow. It's not the easiest way to read, but it's there. So this is
what happened to that paper.

Q: Right. Was the paper also your first connection to the environmental
statistics?

GCT: Yeah, that's right. I got into the environment in 1973. Thank you
for reminding me of that. There's another key motivation for that paper.
See, in 1973 I~got involved in the first ozone project. This is not the
ozone of the stratosphere, but it's the ozone near the ground, ambient
ozone, and this is from Los Angeles. People from Los Angeles called and
wanted our help.

\begin{figure}

\includegraphics{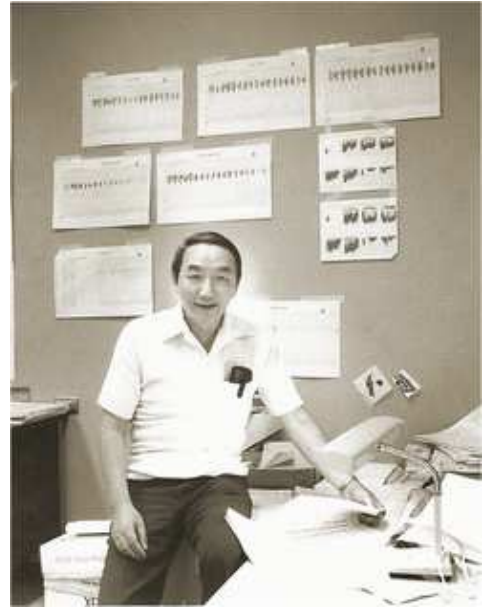}

\caption{Working on Ozone data in early 1970s.}
\end{figure}

What happened was that air pollution there was very bad. The local Air
Pollution Control District set up the monitoring stations and collected
the data. And they got together with industry. The Government had the
data and got together with industry because it wanted to convince the
industry to get together and change or pass the law that regulates the
exhaust/air quality standard.

They wanted somebody to really analyze the data, but they wanted to
exclude all statisticians from California, because those statisticians
might be biased. They wanted somebody from far away who knows nothing
about the problem, which is real and interesting.

There were two reasons they come to Wisconsin. One is the major guy in
the industry graduated from the Chemical Engineering Department in
Wisconsin and so he heard about the Statistics Department. The second
thing is that the Government guy, who passed away a long time ago, and
his name was Hamming (he had a brother famous in numerical analysis),
had some connections with Wisconsin and heard about George Box. I
remember what happened was that they called Wisconsin. Rich Johnson was
the Associate Chairman then. And Rich tried to call Box, but Box was not
around. So he called me. He says, ``George, here's an opportunity and do
we want to do it?'' So I talked to George. He was a little reluctant,
worrying that we might get bad publicity because the money is from the
oil industry. I said, ``Well, why don't I go take a look, talk to them,
and see what happens?'' I didn't want to do it all by myself, so Rich
Johnson and I flew to Los Angeles together and we talked to them. We
were very impressed because these are Government people and then the
industry, they're all scientists, and they just wanted to do a careful
analysis of the data.

\begin{figure}

\includegraphics{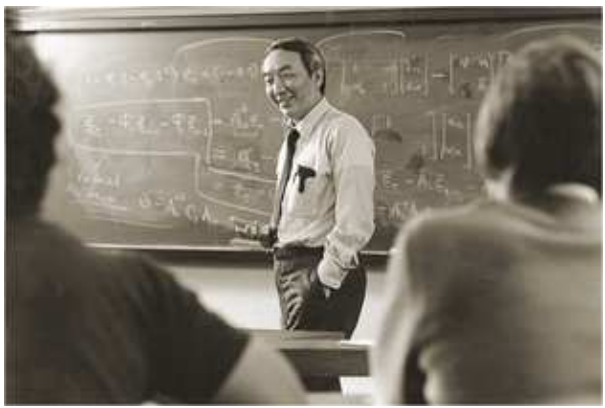}

\caption{Teaching in Wisconsin, in the late 1970s.}
\end{figure}

We were quite convinced and we said we'd do it. And I remember this
thing. As I was leaving, the guy gave me two rolls of tape. I said,
``What is it?'' He said, ``These are all the data.'' I said, ``Are you
going to tell us something about your problem?'' He said, ``No, if we
tell you something, you will be biased'' (laughing). I said, ``No, you
cannot do that. I know nothing about the scientific background of
pollution and so forth. If you don't tell us anything, there's no way we
can be useful.'' And so that's the beginning.

Then they came to Wisconsin and it was a group of about 5 or 6
scientists from the industry and Government. They spent two days
lecturing Box and I and some students to get things started.

Intervention analysis then becomes a natural thing for them, because
they have passed the law and wanted to know the effect of the law. So
that's why in the intervention paper we have pictures of the air
pollution in downtown Los Angeles. (George then showed the interviewers
some color plots of the air-pollution project that were very
informative.)

Q: Could you say saying something about the\break canonical analysis paper?

GCT: My work on canonical analysis started out in the 1960s, I think
probably 1967 or 1968, when I listened to a talk of a colleague, I think
his name is Gill Churchill. He was a colleague in Wisconsin's Business
School and he is a marketing man. He gave a talk about using principal
component analysis to analyze some marketing data. And as I listened to
the talk, I~was thinking that, well, the results sound reasonable, but
one thing we know in all this classical principal component analysis is
that all of the observations are i.i.d. And here you get some business
data, which are clearly not i.i.d. So I was wondering how that would
affect, how the autocorrelations of this would affect, your principal
components. And that was the thing that I had in mind. And so that's
always on my mind and I also looked a little bit, trying to learn
principal component analysis.

So then when I was in England finishing the Bayes book with George, and
also doing things with Cleveland on seasonal adjustment, I remember
talking to George about this principal component problem. That's how the
whole thing started. I remember saying that, well, if it's
nonstationary, is it still reasonable to get principal components? So
then the idea that comes about is to see if you have a lot of series and
they look nonstationary, why do all these series look nonstationary? By
that time I began having some experience looking at business stock
prices and so forth. They all move in tandem. Business indexes often
move in tandem. Now the question becomes: is there some underlying
component which explains that?

Now one way is to do principal component analysis and another one is to
think about transformation and try to explain the relationship with the
past. This gets us to the canonical analysis. That's how this whole
thing developed that way. It's from principal components, a~combination
to explain the underlying and latent fact that explains the
nonstationarity. That idea is very simple, when you have all these
nonstationary things that move in tandem, maybe there are only one or
two underlying components that explain all the growth.

After we came back to Madison, we published the paper in 1977. This took
a long time because our stay in England was 1970. So this and the 1989
paper with Ruey were the two longest papers (laughing). The first
longest paper is the canonical analysis and the second longest paper is
the SCM with Ruey and the third one is the SAR procedure with you
(Pe\~{n}a) (laughing).

We went back to Madison and started to write the paper. The major
difficulty for me is the nonstationarity stuff. I can find various
distributions, but the first thing they assume is stationarity. Then you
can borrow Wold's work and other time-series asymptotics. But the thing
is nonstationary with unit roots and differencing. There was little in
the literature on this topic. I~was trying to see what would happen to
the canonical correlations both in theory and in practice. I know if the
series is nonstationary, the limiting normal distribution breaks down,
but how about when it approaches nonstationarity. It took me a long
time. I remember I spent literally days and weeks trying to understand
what would happen to the canonical correlations when you have roots
approaching the unit circle. I dragged down the paper for quite a lot.
Finally, I thought I understood it. I mean that the proof is basically
right. And later on, I think, Greg Reinsel and others looked through it
and proved it completely.

Once we knew what would happen to the canonical correlations when the
series approach nonstationarity, I thought that we could publish the
paper. But then the surprising thing was the following. I remember in
1977 Box came to my basement in Madison to work on the paper. As we
reworked the Hog example and looked at the components, after you
transformed them, you see one or two is very nonstationary, but we also
see that there are two components very much like noise. First we
dismissed that. All we really wanted are the one nonstationary
components because they explained the growth. But then we looked at two
noise components and played with them for quite a bit and tried various
combinations to explain the components. Finally, we worked out that the
two components, by making some transformation, you get some economic
sense out of them. And after we got that, we say, hey, this is really
interesting. The stable component was a complete surprise, because at
the beginning we didn't work that way. We didn't think that way, either.
We thought it's important to find the thing that underlines the growth.
But then it turns out that the stable components become very interesting
at the end because they have economic meanings.

\begin{figure}[b]

\includegraphics{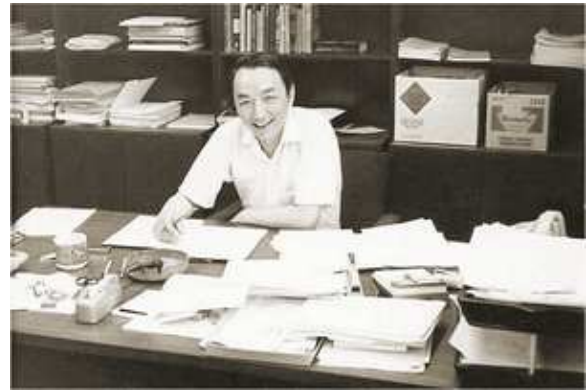}

\caption{Chairman of the department of statistics, Wisconsin, in 1974.}
\end{figure}

Then we began to realize this is very interesting that out of these
nonstationary series linear combinations can be very stable. And I began
to think that, hey, this may apply to all kinds of economic indicators
because they work and move in tandem, although, as we know now, the
problem is still not solved. What is the co-movement and all that stuff?

In summary, you can have combinations of nonstationary series to be
stationary. In the end that becomes the key message coming out of the
paper and out of the example. So that was the history of that. We worked
for about seven or eight years. Oh, from the time the thing hit me 'til
the time we published the paper is about nine or ten years. Many readers
will realize that this phenomenon is the same as cointegration in
econometrics.

I can remember, later on, the work with Ruey in the SCM thing, in the
early 80s, we kept talking about there must be something we can do in
applying the canonical correlation approach with transformation for
model specification. As I remember, it was almost the day before Ruey
left, before we parted in Madison. Ruey called me saying, ``George, this
is interesting.'' At that time he didn't call me George (laughing). You
look at properties of the vector autoregressive model and write the
model in a way that has canonical correlation implications. It is
interesting to use canonical correlation to look at the problem that
way. And that was the key observation that Ruey made and that later on
led to the whole development.

Q: But by that time you had done a lot of work together on this thing
about extended autocorrelation function.

GCT: Oh, yes.

Q: Can you say something about the work together with Ruey?

GCT: Yeah, I said that many times and said it in my letters and now I
can say it again. From day one, I regard this guy, this gentleman, as a
co-equal, although he was my student at the time. And we worked together
really very, very well. There are certain types of students. The one
type is that you give them a problem and you have to still end up
writing the thesis for them. That's one kind (laughing). The second type
is that you suggest a good problem to a student and that student and you
work together. The student does the majority of the work and you keep
providing some insight from experience. That's the way it's supposed to
be anyway. At the end, the student will usually know more about the
subject than the professor, because he or she really spent the time
working on that. And that's the second type. At least the student did
part of the work. And then the best type of a student is he or she wants
to work on a problem. Ruey came to me and said he'd like to work on the
model specification problem, the identification of a mixed model. I
think I'm quoting the truth. I remember I said, ``Yeah, I'm interested
in that.'' So that's how we got to the EACF stuff. Because he wants to
work on that problem and, of course, I happen to be very, very
interested in that too, so it worked out. If you don't have a good
identification tool, the mix model never had a chance. It was not very
much use in practice. It's either AR or MA, because at the time you
could not identify the mixed model. At the beginning, it was always just
ACF, not even the PACF, because PACF is doing regression and it is a
copycat. ACF is very intuitive.

I think I gave him one bit of good advice, which I~take a lot of credit
for. HA! HA! The time we worked out roughly the theory and a lot of
stuff was the time at the end of the Chicago visit, close to the end. He
could have, he had enough to write it up as a thesis. So in~1981 he
could have gone to the market.

Q: In 1981?

GCT: That's right, 1981. He had worked out the basic theory of EACF. At
least we had the procedure and it seemed to work well. Then I told him
that we would go back to Madison. The best thing is to not look for a
job this year. The last thing we still have to work out and let's go
back and spend one more year, then we can get things all sorted out. And
he agreed. Then we went back and in that year I think I probably spent
about three-quarters of the time just with him. That's it, because I had
lots of grants to buy out teaching. The reason I suggest that to him was
there was a prior example, Steve Hillmer. Hillmer graduated and stayed
as a post doc for two years. These two years were really quite
productive. I could pay him as a post doc because of my funding from the
first ground-level ozone project and a grant from the EPA. We not only
published the likelihood paper, which was part of his thesis, but also
worked out all the details of the canonical decomposition stuff for
model-based seasonal adjustment. So I~knew if the person can postpone
his graduation for a year or so and then look for a job, he would be in
much better shape. So that's what I suggested to Ruey and then with him
it worked out really beautifully. By the time he left the Annals paper
was already submitted and the JASA papers were almost about ready.
Anyway, he was a co-equal right from the beginning.

Q: In 1981 you published this very well-known paper with Box about
multivariate time series.

GCT: Right. That paper is interesting. In the late 1970s I started to
teach multiple time series.

Q: Did you use any textbook?

GCT: No, no books. I developed a set of notes. It is interesting that
the course was developed the same time that I developed an applied time
series course in Pittsburgh with George. And I was taking the major part
of the burden in doing that. The reason we could do that is that I had a
great student, Mike Grupe, and the event had to do with the program
development. What happened was that George and I started to teach time
series in the early 70s and the late 60s, because there was a guy by
the name of Bill Ellis at Carnegie Mellon. He marketed advanced
statistic courses. He got George and Stuart Hunter interested in
offering courses on experimental design. And then they got me involved a
little bit; we even had a course on Bayes. Then the next thing to do is
to teach a time series course and I started getting involved with that.
In order to teach time series, you have to have a computer program for
practice.

Let me just give you a little bit of the history of all this. You see,
in school every student would have his or her own ACF program
(laughing). In the late 60s when I was teaching time series, Dave Pack
came and he started to put things together. Still, for a long time
everybody just had an ACF program. There was a nonlinear estimation
routine in Madison's Computer Center, and every student had his or her
own front program to use the nonlinear routine in estimating time series
models. So each time you run one model is a day because the turn around
time is a day (laughing).

Now when you talk about industrial teaching there's no way you can do it
that way. You must have a program. And Dave Pack was the guy who put all
the ACF and PACF together and then started to put a nonlinear estimation
routine together using the NREG program. I guess. Ruey probably still
remember that, because by his time it's all standardized.

Then one of the guys who came to the course in the Carnegie program was
Mike Grupe, and he was working with CompuServe and CompuServe happened
to give us computer time to develop the program. They sent Mike Grupe to
learn this stuff. Mike came and I find he's a remarkable guy; a good
programmer and very interested in going back to school. So I persuaded
him to come to Wisconsin to work a Ph.D. He quit his job and came to
Wisconsin to work with me on time series.

After I came back from Taiwan in 1976, we planned to do the multiple time
series both for school and for industry. But again, the first problem is
the computer program. Generalization of the ACF and PACF, it is not hard
and we got that quickly done. The difficult part is the estimation,
because you need a real good programmer. The theory is not that hard.
You look at it and consider the conditional likelihood based on the
first few observations and stuff like that and then go to the MA part.
At that time I started to get interested in exact MA likelihood with
Hillmer. Mike Grupe was the key guy who programmed the algebraically
very complex multivariate MA likelihood in the late 70s. His program is
still the key element of the estimation program in SCA for multiple time
series. I never had a programmer better than him.

You asked when did I get interested in multiple time series? Well, I got
interested in multiple time series actually through canonical
correlation analysis because that got me interested in general
multivariate data. My background is in economics, economic data are
mostly time series and people are obviously interested in dynamic
relationships among series, and this is multiple time series. So, ever
since 1970 in England I have been involved in multiple time series. I
started to teach multiple time series in 1976 and we developed the
program along the way. In the process we worked on a number of very
interesting examples. So by the late 70s we thought we should write a
paper to introduce the methods along with the package illustrated by
these interesting examples.

One of the major drawbacks of the paper is the problem of identifying
mixed models and we admitted that. And that's ten years later, at least
in my mind. But the good thing about the paper is that the examples show
clearly how this approach works, because at the time one of the
competing approaches is the Granger--Newbold and Box--Hugh approach that
uses pre-whitening. Basically, they're all saying the way you should do
it is that you pre-whiten each series and then work on the residuals. I
had troubles with that right from the beginning because I was in Larry
Hugh's thesis committee.

The first question I asked them is that, you can make this thing very
messy. I showed them by working out simple examples of a bivariate MA.
It is easy to identify bivariate MA models, but if you look at the
pre-whitened series of each one, it's a mess. So I was never convinced
of the Granger--Newbold and Box--Hugh approach. In a way that paper was a
little embarrassing, because basically we'd say that this replaces the
Box--Hugh approach (laughing). Attacking the approach of Granger and
Newbold was one thing, but to dismiss what Box and Larry did was
something else, especially for me at Wisconsin (laughing).

Q: One very important thing about this paper is that it reduces the need
to look at many pictures.

GCT: That's right.

Q: It made the complicated stuff simple to understand, especially with
the invention of those pluses and minuses that are very good.

GCT: I am glad that you raised that. That was one of the proudest
moments (laughing) in my life! Again, it was from examples such as the
hog data. We worked out the correlations and cross-correlations. First
of all, by looking at the matrices, it's a lot better than studying all
the plots of auto and cross-correlations. It becomes relatively simple.
To generalize univariate results to multivariate results, it only takes
a minute. The main problem, however, is about how to comprehend the
results. In fact, the reason that multiple time series and multivariate
ARMA models were not much used is because of the way people identify the
model. To identify a model, you look at all the cross-correlation and
correlation functions. So for two series, you have three things to look
at, for three series you have six things to look at. By that time your
head started spinning (laughing), so for four you give up. If you look
at the early papers, they were all like that. They just have plots and
plots of auto and cross-correlations.

Q: I remember a course that Jenkins gave in which he tried to identify
each element in the matrix. It was not easy.

GCT: Then we thought about looking at matrices. I~was very happy about
that. We're in matrices for two series or three series. It is very
simple if all the elements in the matrices are small. Then you can cut
off and get order of MA models. So that's fine.

When you looked at the hog data, each matrix is 5 by~5. I remember this
was in the basement of Wisconsin, that large room we had there. I
remember that afternoon that Box and I were looking at these matrices.
We realized that even the matrix is not simple, because it has some 25
elements. Then we said, ``Well, which one is big, which one is small?''
So you started to look at things by comparing with standard errors and
crossing out the insignificant ones. Well, we said, ``Basically it will
be just significant or insignificant and we generalize a little bit to
plus, minus and dot.'' Oh, once you see it, that's it. I remember I was
so happy that day. Now we have a good way (laughing). So you might say
that's the only thing that is new in that paper. The examples are very
good examples. Well, the Gas--Furnace example is very embarrassing
because that's the Chapter 11 of Box and Jenkins. It's very hard for him
to put the example in the paper. Basically, it ditches Chapter~11. But
we can do it.

That example convinced both him and me that the vector approach is
worthwhile. But it was hard, because you can see that Chapter 11 of Box
and Jenkins is actually very good theoretically. It solved a lot of
interesting problems about the Coen, Gomme and Kendall by getting to
pre-whitening and so forth.

It is correct to pre-whiten the input and then cross correlate with
output. But it is terrible if you pre-whiten two inputs and then cross
correlate them. It's not the Box and Hugh stuff, it's really Chapter 11
of Box and Jenkins. The 1981 paper didn't replace the transfer function
approach, but the paper comes out with a procedure that's a lot quicker
and can be readily generalized. Once you see that, the adopting of the
multivariate approach to identify transfer function models and so forth
is much more convenient.

Q: After the paper, the theory of integration and co-integration was
developed. The co-integration was developed in economics by Granger and
Engle, although afterward people have recognized that the \textit{Canonical
Analysis} paper used the same idea without naming the term
co-integration. Is your interest in multiple time series trying to find
ways to simplify the dynamic structure and how the idea of canonical
correlation analysis comes into the work of model specification?

GCT: Well, that has a long history too. Remember, that was after we came
here and Ruey came to Chicago to visit frequently. We finished several
papers before we really got to the multivariate analysis. The \textit{JASA} paper
of extended autocorrelation function was first, then we had a univariate
canonical analysis paper in \textit{Biometrika}. I think that's actually an
important paper for understanding the multivariate SCM stuff. (SCM
stands for scalar component model.) As I remember, we had to work all
these out to understand the problem and the procedure clearly.

And on the multivariate thing, as I remember, we did something right
here in this office. There were some key parts that were recognized on
the blackboard here about the SCM stuff. That took a long time to really
understand it. Yeah. That's why it took so long to work out the theory
and then work out a procedure, an integrated procedure, to sort out the
double counting in canonical correlations and in redundant parameters.

Q: Is the sorting concerning the structure of multivariate ARMA models?

GCT: Yes, it is on the multivariate ARMA model and it took a long time.
I still think that the iterative procedure of sorting out true structure
can be simplified. Probably when we start to write a book, that's the
first thing I would like to sort it out.

Q: In other work that you did, during the time at the end of Madison, is
the well-known paper on outliers and outlier detection. Is it the work
that you did with Chang?

GCT: Right. Chang was working on outliers as an assistant.

Q: When did the paper start?

GCT: Oh! The thing started here in 1981. I have three interesting
persons with me. That was Ruey, Ih Chang and then there's this guy
Ahtola working on nonstationarity with complex unit roots. At that time,
Dickey and Harza, both students of Wayne Fuller, worked on unit roots,
but it's one or multiple unit roots, that is, the first and second
difference in time series. The case of complex roots was not done.
Ahtola and I worked on the distribution theory, because Ruey and I did
the consistency of the least squares estimates of all roots on the unit
circle.

But then we got to this nearly nonstationary stuff. We found this very
interesting. There was an interesting paper in \textit{Biometrika} on the nearly
nonstationary case using the score as the statistic. If you look at a
score, you can work out its distribution and then you can see as the
root goes to one, the distribution breaks down. As we worked out the
theory, I thought it is interesting because it's something I can teach
to students. And that's the nice thing about that paper, I thought.
Akaike also likes that paper very much. I remember I talked about it in
Singapore. He liked it and said that this is simple to explain to
students; all the student needs to know is the quadratic forms and
chi-squares distribution and then they can see by their eyes how
interesting it is. So that was with Ahtola.

Q: Which paper is this?

GCT: This is the 1984 \textit{Biometrika} paper on ``Parameter influence about
nearly nonstationary first-order autoregressive models.''

Later on other people worked on this in different ways, such as Wei and
Chan. Ih Chang was working on outliers that were a thorny problem in
developing a program for time series because when you analyze real data
there are always some outliers. The1968 paper on outliers uses a
Bayesian approach, but that was difficult because at that time
computation was difficult so we didn't even think about that approach
for time series. So when Ih Chang was looking for a thesis problem topic
I suggested to her to looking into outliers, and said that this is one
problem we always talk about. She looked at it. The first paper she
studied was the one by Fox. Fox defined two types of outliers, the
additive outlier (AO) and the innovational outlier (IO). But the paper
was a mess.

As I remember, I was looking at the AO and IO of Fox and tried to relate
the stuff. Fox's paper was published in 1972, but I didn't know about it
when I was writing the intervention paper. I think Ih Chang discovered
Fox's paper. At the time the dominating approach for handling outliers
is the Huber stuff. It was quite complicated because it completely goes
away from normality. I like to stay in normality and figure out some way
to test for outliers. The AO and IO idea seemed interesting. As I
remember, it was in Arnold Zellner's study, which was converted from a
garage, that one night I looked at them and realized that we can relate
Fox' approach to the intervention analysis. Once I see that all I have
to do, with AO and IO, is just to simply grind out their estimates from
the residuals using the psi weights and pi weights together. Once you
got that far, the detection procedure becomes obvious. Of course, in her
thesis, Ih Chang did sort out different types of outliers, how to
compare them, the distribution theory and so forth.

Q: Why did the paper take so long to be published?

GCT: Well, she graduated in 1983. And the last part of her thesis was
also quite good, because she also worked out the stuff about
nonstationary series and changes in level. For a given stationary
series, even white noise, if you have a level shift in the middle and
there are a large number of observations before and after the shift,
then the sample autocorrelations are all close to one. We thus
recognized the level shifts and related stuff. But anyway, the thesis
was submitted to \textit{Technometrics} in 1983 for publication. It came back
saying we needed to do some revision. At that time, I was here already
busy developing the business statistics program and courses and she had
a tough time with her job situation. When she graduated she got an offer
to go to work with Professor Der-An Shu in Milwaukee, but she had to
turn it down because she was already apart from her husband a couple of
years. Her husband earned a Ph.D. in Wisconsin and got a very good job
at Kodak and so she had to restrict her job search to around his area
and that was very difficult. Her first job was teaching at a small
college there. She was not happy and moved to Texas working for IMSL for
a while. It was impossible for her to have time to do the reversion and
simulation given her situation and I was busy around here. So the work
was put in the drawer for several years. Also, at the time we published
a summary of the work in a paper with Bill Bell and Steve Hillmer in the
Census Volume. In fact, people were already using the detection
procedure. Ruey got interested and published a paper in \textit{Journal of
Forecasting},
Bill Bell developed a program at the Census Bureau, and the procedure is
also available in the
Scientific Computing Associates (SCA) program. In other words, a lot of
people have been using the procedure, but the paper is never officially
published. It's just in her thesis or a technical report. Finally, when
Chung Chen got his thesis done, I thought that with Chung we might be
able to get the needed simulation done. Chung agreed and did the
simulation. It was quite involved and he spent a great deal of time, so
Ih Chang and I thought that he should be there as a co-author. That's
how it became a three-author paper, and we sent it back to \textit{Technometrics}
in 1988. We almost couldn't get it published, because in a way the
procedure's already known (laughing) and people are using the procedure.
As a matter of fact, the first example we used in the manuscript is the
variety-store data, but we cannot really use it because it was already
in the Census Volume. So finally we used something else.

Q: OK, maybe we can move on. What were the major developments when you
moved from Madison to Chicago in 1982?

GCT: There are two major developments. One is the Ozone Project that I
got involved in and continue to get involved in, and the other is about
the International Chinese Statistical Association (ICSA). I think these
are two major things that probably should be covered. Which one would
you like me to talk about first?

Q: Any one is fine.

GCT: Well, I mean what are the specific questions that you want me to
answer?

Q: Perhaps you can talk about the ozone project.

\section{The Ozone Project}

GCT: What happened was that Bill Hill talked to me about an ozone
project in 1978 or 1979, I think it was 1978 when he first called me. He
had worked on the stratospheric ozone and the project was getting a lot
of press and became an important public issue. How does the CFC relate
to the depletion of ozone in the stratosphere? At the time I was just
getting out of the Los Angeles ozone project, the ground level ozone and
air pollutant stuff and was also involved with all the Census Bureau
stuff, so I told him I could not really do it, because the Census things
already costing me quite a bit of time. So he went away and a year later
he came back again. He said he really need my help in the sense that he
can do time series analysis, trend analysis, intervention analysis and
all that, but he worked in the industry and was concerned about the
credibility problem. He needs some sort of academics to really do good
analysis and make the results credible. A very important decision will
be made because if you stop the use of CFC, it would change a lot of
stuff we use like refrigeration and the coolant in air conditioning.

So I got the project and said OK I'll give it a try. I was first looking
for somebody to work with me on that. Instead of finding some post
doctors, I looked for a younger faculty. I talked to Jeff Wu and he
thought that was a bit too much, too big a project to kill his time. And
the second person I tried to persuade was Greg Reinsel, who was a new
faculty in Madison teaching 709-Mathematical Statistics. He is a time
series guy and he knows all the ARMA models. So I tried to persuade him
that this would be a great opportunity for him to look at real data. He
agreed. Then we had a post doc from Taiwan and Doug Nichyka (who became
a very famous statistician at North Carolina State and now at NCAR) and
another student, Rich Lewis. So with two or three students, we got
together to start the project. That's how thing began. At the beginning
the project was supported by the CMA, the Chemical Manufactures
Association. That was Bill Hill. He worked with us. Then pretty soon
Lane Bishop also joined the industrial team.

We met with industry people and people from NASA, EPA, and NOAA and
Canadian Environmental Service regularly once every three or four
months. This gradually develops into a team of atmospheric scientists,
modelers and statisticians working on the atmospheric ozone and later
temperature problems. They called it a Tiger Team. Basically, if we want
to really understand the problems, scientists need statisticians and we
need their scientific input. Over the last two and a half decades the
team has published over 30 papers, most in top rank geophysical science
journals, and have achieved quite some impact in the scientific
community.

\section{ICSA and Statistica Sinica}

Q: You are deeply involved in the International Chinese Statistical
Association and are the
Founding Chair-Editor of the journal \textit{Statistica Sinica}. Can you mention
briefly about ICSA and the journal?

GCT: The origins of the ICSA and the journal \textit{Statistica
Sinica} are somewhat intertwined. I first attended the ASA annual
meeting in 1961 and, to my surprise, there were only three Chinese
statistical experts, two in economics and one in biostatistics. Things
started to change in the late 60s and early 70s as increasingly larger
numbers of Chinese/\break Taiwanese math. students came to study statistics in
the US. We bought a house with a basement in Madison in 1967 and
started to invite Chinese statistics students and their families to the
Thanksgiving dinner in the basement. This ``dinner festivity'' grew to
more than 80 participants and lasted for more than 20 years. Many
students came to prepare the food the day before Thanksgiving and help
cook dinner the next day. We have very nice memories about these
dinners. As a matter of fact, many former students know my wife much
better than me because of the dinner and opportunities to get together.
It was at the 1968 dinner that I realized we perhaps need an
organization to promote communication and collaboration among Chinese
statisticians. We started with an informal association called the
Chinese Statistical Society in US. With the help of 8--10
enthusiastic volunteering students, a hand-written bulletin that
contains the directory of Chinese statisticians was published in the
following year. The principal student leaders for the first two years
were Austin Lee and Der-An Hsu. After the third year I thought that it
is better for the Society to broaden its base by rotating it around the
States. I asked Professor Y. S. Chow of Columbia University for help. He
recruited Min-Te Chao, later the Founding Director of the Institute of
Statistical Science, Academia Sinica, to be responsible for the
administrative activities of the Society. At that time, Chao was at Bell
Lab. Without the help of graduate students, the publication of the
annual bulletin quickly became an impossible burden for any single
individual. Three years later, the Society moved back to Madison and was
renamed Chinese Statistical Association in America.

Besides the bulletin, Chinese statisticians got together to have dinner
each year at the annual ASA meeting as a means to get acquainted with
one another. As I recall, this started in the early 1970s at the St.
Louis meeting with Hubert Chen as the first organizer. It began with 10--20
people, but grew quickly to more than 100 by the beginning of the
1980s. It also became a regular event with an informal meeting in the
late afternoon of Wednesday followed by the dinner. With the expansion,
we started to wonder whether such an arrangement is effective and
sufficient for promoting the communication among Chinese statisticians.
The final push to have a formal association has to do with \textit{Statistica
Sinica}.

\begin{figure}[b]

\includegraphics{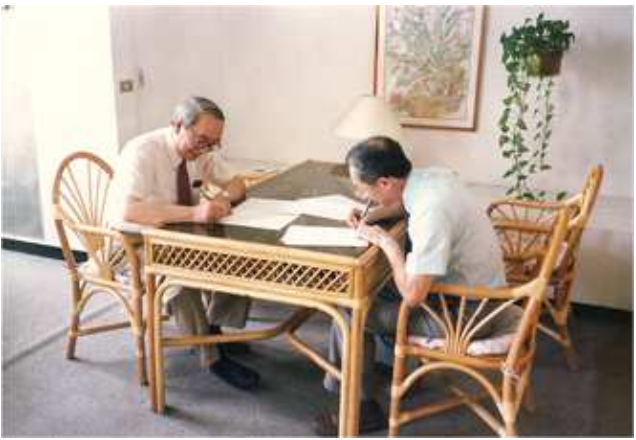}

  \caption{Signing agreement of Statistica Sinica in 1987.}
\end{figure}

Three key developments occurred in 1986. First, I~returned to Taiwan for
the Academia Sinica members meeting and had a chance to meet with
Director Chao and other statisticians. Chao suggested that the Institute
is sufficiently mature and has budget to launch a new statistical
journal. We felt that to publish a new journal, it is best to involve
all Chinese statisticians inside and outside of Taiwan. Second, the 1986
ASA meeting was held in Chicago and Jia-Yeong Tsay suggested at the
afternoon meeting that the time is ripe for the association to be
formalized. He, Grace Yang and Gordon Lan formed a committee to draft
the constitution of ICSA in a spirit similar to the ASA Charter. Third,
at the dinner James Fu told me of his plan to launch a new statistical
journal. I suggested to James and Min-Te that it seems better to combine
their efforts to establish a world class journal. They came back to me
about the Christmas time that year saying that they decided to cooperate
and wanted me to formally launch the journal. To make the long story
short, I consulted with several senior Chinese scholars, including the
late Professor Shein Ming Wu from Madison. They were all very supportive
and gave me valuable suggestions, including having a strong local
support at Chicago. As you know, Wing Wong was in Chicago then and he
gave me his whole-hearted enthusiastic support. Ruey Tsay and Xiao-Li
Meng also came to Chicago shortly after. Furthermore, I also obtained
enthusiastic support from Smiley Cheng, T. L. Lai, L. J. Wei and Jeff
Wu. Thus, I~decided to accept the challenge. An editorial board was
formed in April 1987 and I served as the Chair-Editor. Because the
Institute cannot sign an agreement with an informal association, the
ICSA was formally established in 1987 to jointly sponsor the journal
named \textit{Statistica Sinica}. I was elected as the first president of ICSA
for a one year term to help get it started.

Over the last twenty years, ICSA has grown to become one of the largest
statistical organizations in the world. It has not only provided
services and communication among Chinese statisticians in North America,
but also helped promote statistical theory, application and education
among Chinese communities in mainland China, Hong Kong, Singapore and
Taiwan. To this end, it has held international conferences in Hong Kong,
Taipei, Beijing, Kunming and Singapore. Another success story of ICSA is
the annual \textit{Applied Statistical Symposium}. Many members of ICSA are
biostatisticians working in the pharmaceutical industry and in the
federal agencies such as FDA and NIH. Jia-Yeong Tsay and Gordon Lan
initiated and organized a successful half a day meeting called \textit{ICSA
Biopharmaceutical Statistics Symposium} in Washington, DC in 1990, and it
was expanded to include other areas beside biostatistics and changed
into the current name two years later. This symposium has become an
annual event and grown into a three-day affair attracting more than 200
participants from around the world.

The twenty year history of \textit{Statistica Sinica} is equally lustrous. The
first issue, appeared in 1991, was immediately recognized and highly
praised, and the journal has been widely supported by colleagues and now
is generally regarded as a top ranked journal in our profession. Much of
the credit has to go to a stream of prominent chair editors including
Jeff Wu, C. S. Cheng, K. C. Li, Jane Wang, X. L. Meng and currently Peter
Hall.

\section{Making Statistics More Effective in Business Schools
Conference}

Q: Tell us about the Making Statistics More Effective in Business
Schools Conference.

GCT: What happened was this. In the mid 1980s, on the one hand, there
was this quality movement that began to happen. It attracted a lot of
attention because at the time the American industry was going through a
pretty tough period. Got beaten up by the Japanese on manufacturing and
all this. Also, at the same time, the statistical profession started to
really expand and to get into many different areas. A particularly
important area is engineering. In order to develop effective courses in
engineering, you have to work with engineers and Bob Hogg at the time\vadjust{\goodbreak}
was pioneering this effort in Iowa. He organized a conference, a lot of
people attended and focused in on how do we offer more, better courses
for engineers.

I remember, in 1985 I was invited to New York to attend one of the
quality conferences. On my way there I was thinking that now with a good
program going in Chicago Business School, we should get together with
other teachers of statistics in business schools because we're the
largest school, and have a forum to discuss our common experiences and
what kind of problems we have. So on the way to New York I was thinking
about all this on the plane and I said, well, maybe I should consult
with Bob Hogg and see what we should do. It was quite interesting
because when I~got to the conference site and saw Bob, before I opened
my mouth, he said, ``George, I want to talk to you.'' I~said, ``Well,
there's something I want to ask you too.'' When we got together he was
talking to me about his conference and things. He said that he was doing
that and thought that somebody should do that with business and
economics. He was talking about me. And I said that's exactly the kind
of thing that I thought would be good to do. He said, ``Well, I'll help
you with that.''

When I came back I talked to Harry Roberts about this idea. Harry has
been my mentor in Chicago GSB and I've known him for many years. In
fact, when I graduated the first time I came down to Chicago, I went to
visit with him and gave him a copy of my thesis.

Q: When did you meet Harry Roberts?

GCT: In 1962 and he has been always very encouraging to me ever since.
Anyway, when I came down he was a senior guy and had a tremendous
reputation for his contributions to the school. Whenever I have any
plans or any ideas, he's the first one I'll consult with. And he's
always been very encouraging and helpful, as also Al Madansky.

So I talked to Harry. I said, well, this is something that seems
worthwhile doing. Would you be happy to do it together? He immediately
thought it was a great thing. In addition to Harry and I, we need a
young person, and George Easton agreed and the three of us got together.

In order to organize this conference, we first of all need some support
from the school. So I went down to the Dean's office to get help. Harry
Davis was the Deputy Dean. He was very helpful and promised money. We
thought that we want to do it, not just in Chicago, but each year at a
different place. First of all, it's pretty tiring to do every year, and
second thing is that it's not very helpful. The helpful thing would be
to go to different places and have this forum in different places and
different universities. But, on the other hand, at the beginning I don't
know how many will participate and that we might have to do it more than
once. Then Harry Davis said, ``How often?'' I said, ``Well, maybe five
years and then we can probably get us established.'' So he said, ``Well,
George, OK, we'll support you for five years and give you money to
support this three to five years.''

So with all of that, we got together and there were two things. One was
to develop a directory and Harry Roberts took the initiative to go
through the ASA directory, literally page after page, to find out from
the directory who are teaching in the business schools. Just from
people's locations and addresses, he literally did it that way
(laughing). In the first year we really did a grand job. We covered all
the application fields in the business world, plus quality and industry.
We had to develop connections with business and also with our colleagues
in the function area of the school. So we'd contact people in finance,
accounting, marketing, management, quality and industry. These are all
the elements of that two-day meeting. 
\begin{figure*}

\includegraphics{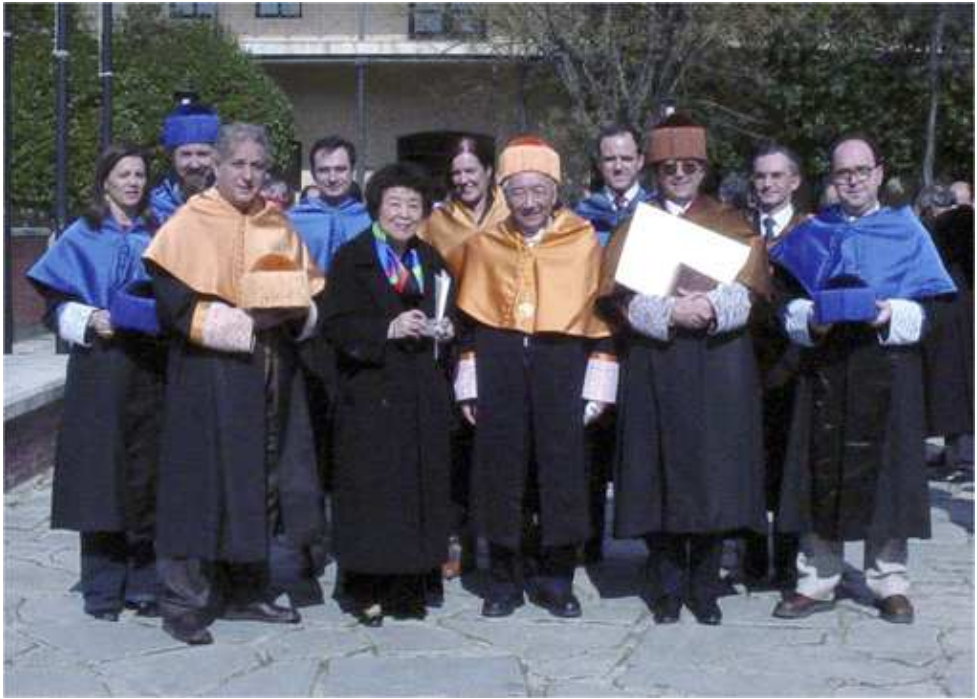}

  \caption{With Barbara, Daniel Pena and other faculty members of University Carlos III in 2003.}
\end{figure*}

We got the biggest classroom we had. At that time we had about 130
seats. So we kept worrying about how do we fill the 130 seats. And we
sent announcements to everywhere we could find out and got a school that
got a list of other business schools. Now the school gave us a few
thousand dollars to organize the conference. Did you know what happened?
It's just totally incredible, because we filled up the room, every seat.
We got people from 66 institutions in America and outside, for example,
like Hong Kong. And there was somebody from Europe too. We also got a
very good business representation, even from some large companies. We
organized a dean's session. Jack Gould was the Dean at the time and then
we had the Wisconsin's Dean and I think Stanford's Associate Dean or
someone like that. And we also got people like Schlaiffer and John Pratt
from the Harvard Business School. This was in 1986, and I think the
second one was in New York at NYU.

Q: Was it in 1987 or later?

GCT: Every year. So far we've never had to do it again. There are always
people wanting to do it. The second year was in New York and the third
year in Madison, Wisconsin. Then one year it was in Ann Arbor and so
forth going around. The first few years were very, very successful. In
the conference we had two types of sessions. One type is industry and
area specific, and the second is teaching. How do you teach a basic
course? That was where the tradition continues. Of course, the first
conference was the most successful because we covered almost all fields,
and it's the beginning. After that we have an area of concentration
every year, but teaching is always an important element of that
conference. That's sort of continued down to present time.

Q: It's been running for 15 years now. What type of impact has it had?

GCT: One of the major impacts is on the textbooks. Many of the textbooks
that came out for business statistics in the early 90s all acknowledge
the inspiration they experienced in these conferences. So it has a major
impact on textbooks. Also, for a lot of people their teaching has
greatly benefited by their going to the conference, and sharing
experiences. That's probably the major impact.

\section{Time Series Conferences}

Q: Another conference that you have been very much involved with is the
time series conference.

GCT: The time series conference was like this. Arnold Zellner is the one
who established the Bayes conference in 1970. It runs twice a year and
it is still going. I think about 1977 Arnold and Bill Wecker got
together to do one in time series. I remember I came down to Chicago to
participate. In 1978 we did that in Madison. In 1979 it was down in the
Census Bureau in Washington, DC, and it disappeared in 1980. When I
came to Chicago in 1982, the first thing I did was talk with Bill
Wecker. He's a great guy. I said, well why don't we get together to
revive this? So Bill got the money from NSF through NBER to finance it
for five years and we organized the meeting in Chicago in 1983. It was
quite successful.

Then Bill Wecker left Chicago. When he left, I~said, well, we organized
and worked so well together,\break  I~would certainly hope you can continue to
participate and keep this going. And he promised, so we divided the job
such that he kept the money coming in and I am responsible to locate a
place each year. We went to all sorts of interesting places, Davis, San
Diego, Carnegie Mellon University, Madrid, Vienna and Taiwan and all
that. Let me just say this though. That was one of the proudest things I
was associated with because this annual conference really mixes time
series statisticians and econometricians together. The conference is
always very well attended by excellent people.

\section{Future}

Q: Maybe we can move to talk about more general things. We'd like to
know how you see the future of statistics as a profession.\looseness=1

GCT: I'm basically an enthusiastic guy. I do think that the statistics
profession has a great future, if we do it right. And the reason is that
the whole world is getting more and more quantitative. In business and
in science, no matter what field you're in, it depends more and more
upon quantitative information and then how do you collect them? How do
you analyze them? I'm very enthusiastic because you can see the need for
statistics all over in biological sciences, in natural sciences, in
physical sciences, in environment and meteorology and, of course, in
business and economics. Every field in business, in marketing, in
production, in finance, I don't think they can operate without
statistics. So the use of statistics has become more and more
widespread, and the need for statistics is just increasing, there is no
end. On the other hand, if we don't rise to the challenge, then other
people gradually will take over our functions. An example is data
mining. A problem could be the training of pure mathematical statistics;
it's a long story about this. The feeling I have is that if you look at
the growth of our profession, we really don't grow that much, right? I
don't think the size of the ASA is more than 20,000. Is that right?

Q: Slightly less, about 17,000.

GCT: Close to 20K. But I remember even 10 or 15 years ago when Barbara
Bailer was the President, the number was about 15,000. This was a long
time ago, and we didn't grow that much. Very few places have
undergraduate statistics program. And if you look at the graduate
program, the Ph.D. is, of course, naturally not that big. But if you
look at a Master program, it's not big, either. Our students are very
bright, but when they go out to work, they're at a disadvantage because
they don't know any application areas. A few years ago I thought we
should change our program and make all students to have one area of
application. This was basically done in Carnegie Mellon, in the
statistics program, but the most successful place is Columbia. They come
out with the same idea independently and the Statistics Master program
consists of 8 courses or something like that. You can do it in one year
plus two summers. Out of these 8 courses, only 5 are on statistics
methods and theory, the other three are concentrated in their
application area. This program saved them, from less than 10 students in
three years they got up to about 70 students, and now it is well
established.
\begin{figure}

\includegraphics{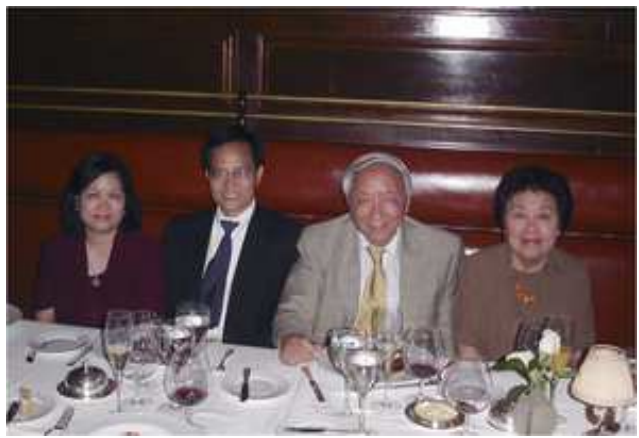}

  \caption{With Tsays in 2003.}
\end{figure}
\begin{figure*}[t]

\includegraphics{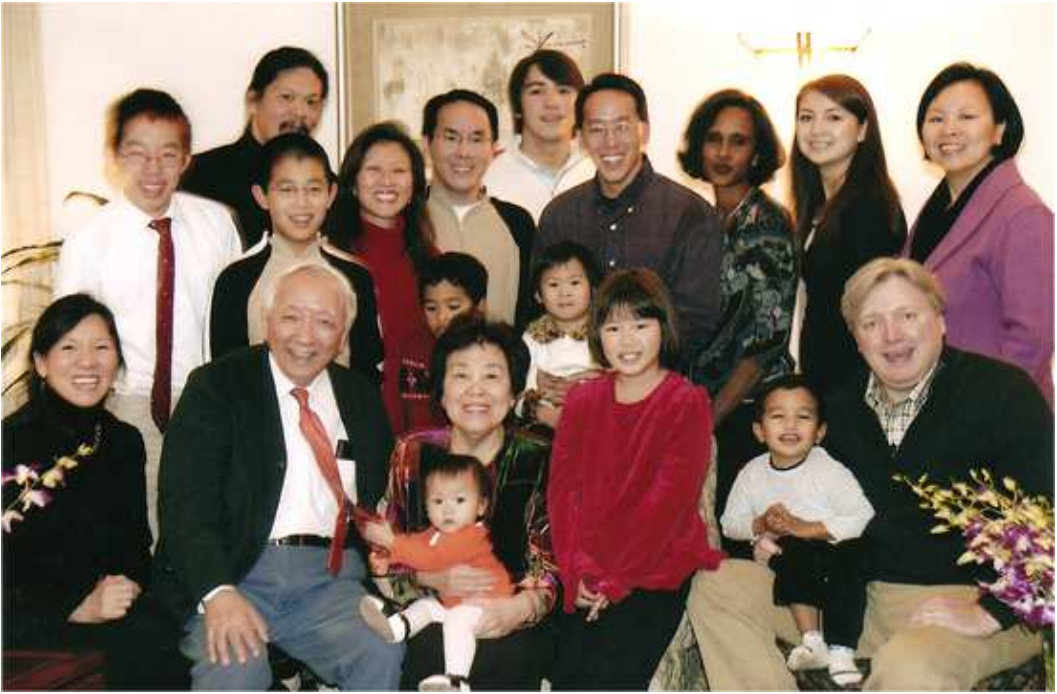}

  \caption{The Tiao Family in 2007.}
\end{figure*}

So I think it's useful to establish some kind of joint Master program
with an area of concentration in application. My background is in
economics, but when I get involved in environmental data my application
background in economics actually helped me to get into other areas. So
that way we can probably expand our Master program.\looseness=1

Q: What are your plans for the future?

GCT: I think that what I'd like to do for the rest of my working career
is to go back to spend more time doing research, because I think in the
last ten years or so I have spent way too much time developing programs.
For instance, I have been working with Ruey Tsay and Rong Chen to
establish in Taiwan and Beijing a quantitative finance\vadjust{\goodbreak} department with a
program in which students have a basic good training in statistics and
then get into this important area.

Q: What kinds of things do you like to do and enjoy when you are not
working.

\begin{figure}[b]

\includegraphics{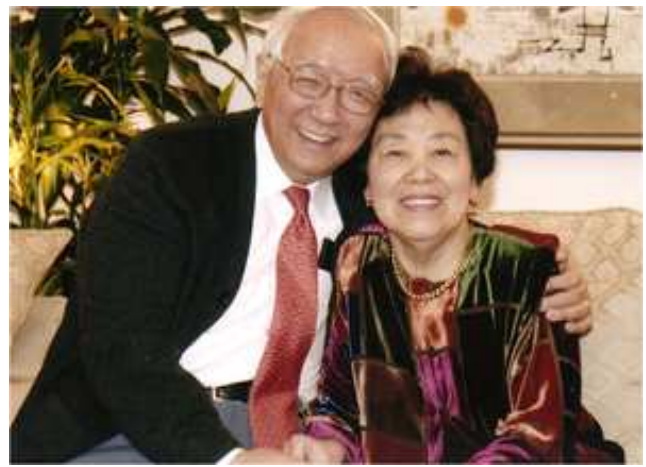}

  \caption{With Barbara in 2007.}
\end{figure}\

GCT: Oh, basically I have a very good family. I have a wonderful wife.
We've known each other more than 50 years now (laughing), which is
rather unusual these days.

Q: And your wife, Barbara, she's very much involved with music, right?

GCT: She is very much involved with music and she sort of gave up her
music career in the early years when we had to go through graduate
school and then the first few years because we had kids quite early. And
we have four kids. They are all grown up now. They are doing well. So I
have a very, very wonderful family life.

Q: How many grandchildren now?

GCT: There are five now. Hopefully there will be more to come. So I am
enjoying the third generation; it is a great joy.\vadjust{\goodbreak}

When I was young I used to play a lot of bridge and also learned to play
golf. But in the last 30 years I have basically given it all up. I was
fortunate, because I was really brought up in one culture and then
adopted another culture. I still go back to China and Taiwan very often.
So I continually kind of like live in both, both are part of my life.
When I go to China or Taiwan I~have a feeling that I've never left
America, and when I come back to America, I feel that I have never left
Taiwan or China. I don't play any music, but I enjoy music and also
enjoy art and literature, from both cultures. So I think that I have
lived a very fruitful life (laughing) and am looking forward to many
more years (laughing) hopefully.

Q: OK, very good. Is there anything that you would like to add?

GCT: I just want to thank both of you for listening to all this.

Q: OK, thank you very much George, and we all miss Barbara who passed away in March 2008.\vspace*{-12pt}

\end{document}